# HOW MANY PARAMETERS TO MODEL STATES OF MIND?

Krzysztof Kułakowski, Piotr Gronek, Antoni Dydejczyk

Faculty of Physics and Applied Computer Science,
AGH University of Science and Technology,
al. Mickiewicza 30, 30-059 Cracow, Poland
E-mail: kulakowski@fis.agh.edu.pl

**ABSTRACT**

A series of examples of computational models is provided, where the model aim is to interpret numerical results in terms of internal states of agents' minds. Two opposite strategies or research can be distinguished in the literature. First is to reproduce the richness and complexity of real world as faithfully as possible, second is to apply simple assumptions and check the results in depth. As a rule, the results of the latter method agree only qualitatively with some stylized facts. The price we pay for more detailed predictions within the former method is that consequences of the rich set of underlying assumptions remain unchecked. Here we argue that for computational reasons, complex models with many parameters are less suitable.

## INTRODUCTION

Since the times of Max Weber, social scientists agree that mere external observations do not provide sufficient information to understand human actions. Beliefs, expectations, emotions and norms play a major role there and should enter to theories and models. On the other hand, these internal properties cannot be measured directly; information on these variables can be collected by interviews, and the relation between the interview results and the internal properties remains at best subtle.

The aim of this text is to highlight the connection between emotional and behavioral aspects in selected computational models of states of mind. As a rule, these models are postulated with the aim to interpret some observed effects; a successful model should also suggest the way to refine methods of research. If a model does not offer any contact with reality, it deserves to be described with the Pauli's famous statement "not even wrong".

It is obvious that a model with more parameters allows to reproduce observed data with more details. In our view, this option is a dangerous temptation; we argue that, paradoxically, models with more internal parameters are less suitable to infer about internal variables. Our argumentation is directed against models, which intend to capture reality by fitting parameters. This procedure is well established in physics; also in mathematics we sometimes determine unknown parameters from the condition of existence of a nontrivial solution. On the contrary, in social sciences parameters can be fixed only rarely. Even if so, their values fluctuate from one society to another, from one time instant to another. It is not sufficient, therefore, to demonstrate that this or that social effect can be reproduced within a given model. We should check how our model works for the parameter values taken from some ranges, justified by the model context; only then we keep control of the model outcome. The procedure, known as the sensitivity analysis, is hard to be applied for purely computational reasons if the number $K$ of parameters is large; the number of program runs increases exponentially with $K$. Consequently, the analysis of the results is more likely to remain superficial.

## MESSAGE RECEIPT

The first story to bring up here is a method proposed to unify two earlier approaches of public opinion (Zaller 1992; Deffuant et al. 2000; Kulakowski 2009). Actually, both models (Zaller 1992; Deffuant et al. 2000) readily apply to the more general problem of social communication. The Zaller book provides an extended frame of analysis of social receipt of messages, as dependent on their content and on individual characteristics of recipients. At the core of this frame, a multidimensional parameterization was developed, with separate parameters related to the credibility of the message, the awareness on resistance to persuasion, the predisposition on resistance to persuasion, the message intensity, the strength of a relationship between awareness and reception and some others. This kind of modeling aims to refer to individual emotional predispositions of the message recipient, related to particular messages. As the outcome of the mathematical formulas, we get the probabilities that the recipient receives and accepts a given message or simply ignores it. Then, these probabilities allow to infer on the recipient's behavior: a reaction triggered by the message or lack of it.

A general problem encountered with multidimensional parameterizations is that the sensitivity analysis cannot be performed. With ten parameters and only three trial values of each (large, small, medium) there is already almost 60 thousands model results to be analyzed. What is even more important here, the number of parameters should be smaller than the number of calculated outcomes; otherwise the model is reduced to a parameterization, which allows to encode potentially each expected/demanded result in the input. More about principles of social modeling can be found in (Edmonds 2000; Edmonds 2005; Helbing and Balietti 2011).

The Deffuant model and its later extension seem to be a remedy to the problem of many parameters. The model (Deffuant et al. 2000) explores the concept termed 'bounded confidence', what – in a broad sense - means that people ignore opinions, messages and other persons which and who are more distant, than some prescribed threshold. In this way, the idea of distance is introduced to psychological and social considerations. This idea, although it can seem trivial for a physically oriented mind, brings two specific properties: first, it is expressed in numbers, and second, it fulfills the so-called triangle inequality. The latter means that the distance between two objects, say A and B, is not greater than the distance between A and C plus the distance between B and C; this should be true for any object C. This inequality, basic for Euclidean geometry, has never been proved in relations to social sciences; on the contrary, it is possible to break it, what is known as intransitive preference (Noteboom 1984). It appears that the introduction of distance is a strong condition which allows to reduce the number of parameters to one or two. Although as arbitrary as the Zaller parameterization, it makes the model of social communication much simpler.

In the unified version (Kulakowski 2009) of the Zaller-Deffuant model, points in a planar area represented messages on two basic issues, say safety and welfare. Agents' knowledge and experience with respect to these issues were growing, as the agents were able to receive messages which were not too far from messages received previously. The threshold distance from a previously received message to a newly accepted, although far, one, represented the mental ability of an agent in the unified model. The basic result of (Kulakowski 2009; Malarz et al. 2011; Malarz and Kulakowski, 2012) is that agents with smaller abilities are more prone to extreme opinions. Apparently, this result does not depend on the choice of the threshold value. In this way, the difficulty of the multidimensional parameterization is evaded.

Above we noted that the Zaller model allows to infer about the behavior of the message recipients; they react or ignore the message, depending on the relation between the message and their ability and experience. The same is true with respect of the Zaller-Deffuant model. A trivial example is when the warning about an emergency is announced in an unknown language; this warning will be ignored till the moment when unquiet behavior of local groups will be imitated by strangers. Less trivial is the conduct of children when alarm is heard in a school building; this signal will be probably interpreted by them with less attention than by teachers, who are responsible for their evacuation.

## WHOM WE LIKE, WHOM WE FIGHT

Another approach to be mentioned here deals with the problem of (Kulakowski and Gawronski 2009). It is motivated by the Prisoner's Dilemma, but is free from the parameters which describe payoffs. In this model, the probability that X cooperates with Y depends on the reputation of Y and the overall propensity (altruism) of X to cooperate. In various model variants, reputations and altruisms of agents vary or remain constant. Recently we could demonstrate, that an exclusion of agents with bad reputation does not undermine the social rate of cooperation (Jarynowski et al. 2012). Here again, an interpretation that cooperation is often limited to people with similar social status, seems natural (Weber 1978, p.932). We can make this conclusion more firm, using the concept of reciprocity (Fehr and Gachter 2000) ; people who do much better will not reciprocate my cooperation, because they are not afraid of my punishment; I will not reciprocate the cooperation of a poor for the same reason. Also, the numerical outcome can in principle be verified by a careful experiment.

In this model, the connection between behavior and emotions is assured by the link between a cooperative behavior, a hope for reciprocity and the fear of punishment. The results reported in (Jarynowski et al. 2012) indicate, that once these emotions are absent, cooperation fails. A historical example is provided by the evacuation of American diplomats from Saigon in 1975 (McNamara 1997), when the evacuees could not gain anything by cooperation with their Vietnamese allies.

The description of the next model should be started from a reference to experiment. In 70's, Wayne Zachary investigated social relationships among 34 members of a karate club at an American university (Zachary 1977). Zachary wrote down these relationships as a 34x34 connectivity matrix, indicating who had contacts with whom. During this research, a conflict appeared between the administrator and the teacher, and the club happened to divide into two groups. The matrix and the actual division (who with whom) entered to a data base, useful for social analysts. In particular, a simple set of nonlinear differential equations has been proposed to describe the time evolution of relations between agents (Kulakowski et al. 2005). The calculations – with the connectivity matrix as an input – exactly reproduced the division of the club.

The driving mechanism was the attitude to remove the cognitive dissonance (Festinger 1957); an emotional discomfort which we feel when some parts of our environment are mutually incoherent. In the case of the Zachary measurement, the discomfort experienced by the club members was related to their colleagues; some seemed to be more sympathetic, some less. These classifications were not mutually independent: the karatekas ordered their views according to the principle "friend of my enemy is my enemy" and the like. As a behavioral consequence, the club split appeared to be in accordance with their internal feelings.

The examples given above provide an evidence that the inference from/to emotions and behavior is possible at the model level. In all these examples some kind of behavior (reaction for a message, cooperation, solidarity) was one of two options, and the adherence to this behavior was motivated by a definite mental state which also could appear or not appear. We note that in some social situations, a given emotion can be believed to appear without alternatives. In this case, the aim of modeling is just to investigate consequences. A good example is the text (Malarz et al. 2006) on the Bonabeau model (Bonabeau et al. 1995); the latter was formulated with a reference to animal rather than human societies. Most briefly, the problem can be presented as follows. A group of agents wanders a given area, and those who met have to fight. The fight outcome is that the winner gets some goods from the loser. Also, the probability that an agent wins depends on his amount of goods before the fight. On the other hand, the differences between wealth of agents decrease between fights. The problem is, if the variance of wealth will be large or moderate? A phase transition between those two options was previously identified in the literature, and our text (Malarz et al. 2006) is devoted to an analysis of this phase transition. Perhaps the model could be an illustration of increasing differences in power between local rulers in medieval Europe. A beautiful and deep description of this process was given by Norbert Elias in his monumental book „The Civilization Process" (Elias 1939). However, we do not learn anything on human beings from the numerical results. The driving emotions, supposedly fear and hate between rivals, is built into the model without alternatives.

**FEELINGS IN CROWD**

When looking from this perspective, there is some analogy between the Bonabeau model and the modeling of crowd dynamics, as in the so-called Social Force Model (Helbing et al. 2000). There, pedestrians are represented as particles in a two-dimensional space, with appropriately chosen masses, radii, elastic and friction coefficients. We note that this careful design allowed to reach numerous interesting properties of the crowd, with undoubted accordance with reality (Johansson et al. 2007). The human nature of the simulated pedestrians manifests in that they prefer to keep mutual distance (hence „Social Force") and in their ability to select direction and velocity of their motion. This characteristics can be enriched by an individual modification of their parameters or even by some manipulation of their purposes (Gawronski and Kulakowski 2011; Gawronski et al. 2012), but all that is to be done by hand.

This list of our social modeling is to be compared with the research strategy applied in large scale by Treur and Sharpanskykh and their cooperators (Sharpanskykh 2010; Bosse et al. 2011). The declared aim of the paper (Sharpanskykh 2010) is to define relations between different cognitive processes of an agent in a socio-technical system. The list of these processes is derived from the literature. These are: belief revision, trust dynamics, generation and development of feelings and emotions, and decision making. Reading the text, we learn that the essence of work is to introduce the desired dependences of related variables by properly placed instructions of a specially designed computer language. In a section „Experiments", three runs of the simulation of an evacuation are reported. As numerical results, three different curves are presented on the time dependence of the number of persons in the room.

The paper (Bosse et al. 2011) reports a more direct connection to experimental data. Namely, the aim here is to reproduce the motion of people, filmed during the panic outbreak in Amsterdam, May 4, 2010. To achieve this, a total difference between filmed and calculated trajectories was minimized, tuning two global parameters and individual time dependences of the maximal speeds of people involved. This difference – a measure of the simulation error – was compared for three different cases: people did exchange emotions, people did not exchange emotions, people did not move at all. The exchange of emotions was built in to the general simulation frame, the same as in the previous paper (Sharpanskykh 2010). The first option gave the smallest error. In their conclusions, the authors underline this result as an argument that people do exchange emotions.

We are tempted to suspect that it is the general, multidimensional frame used here what makes the contact with experimental data superficial. The conclusion that panicking people do exchange emotions is certainly reasonable. However, one can ask if this conclusion could not be obtained within a simpler model? Similarly to the Zaller approach, the general modeling frame used by Sharpanskykh and Treur does contain so many internal parameters of the simulated agents, that the abundance of these parameters disables any systematic analysis of their role.

## DISCUSSION

Still, the issue becomes more complex when we realize that the multidimensional modeling could be defended as follows: if we fix all parameters but one, the obtained model should be formally equivalent to a model with one parameter. What is wrong with adding new parameters if we keep them constant? Perhaps a subtle but simple answer can be found in the prescription of modeling, given in (Edmonds and Paolucci 2012) in a book review. The authors write: „To assess the usefulness of a modeling technique one has to look at the strength of three stages in the use of a model (...): (encoding) the map from the known or hypothesized facts and processes into the model set-up, (inference) the deduction of results from the set-up to the outcomes, and finally (decoding) the mapping of the results back to the phenomena of concern. Roughly, the usefulness of a model is the reliability of the whole modeling chain: encoding + inference + decoding." The answer could be that in multidimensional models, this reliability is particularly difficult to be controlled.

The models brought up here as examples are different. The Zaller model of public opinion does not provide more insight, than the data it uses as an input; this is just a translation from the data to a set of coefficients, which can be measured only through these data. The bounded confidence model brings instead the concept of distance and one control parameter. One can wonder, if messages can be distributed in a geometrical space or rather on a network. Basically, we should be able to verify the Deffuant model by checking if the small world effect applies to the set of opinions; we imagine that this could be done by carefully designed interviews. Also, if the triangle inequality is broken in a given system, the whole concept of distance cannot be maintained. Taking this into account, we admit that the model assumptions are susceptible to a falsification; in terms of Wolfgang Pauli, this model can be right or wrong.

In a reformulation of the Prisoner's Dilemma multi-agent game (Kulakowski and Gawronski 2009) a distribution of reputations of $N$ agents about $N$ others has been used. The essence of the model was to propose the rule of evolution of these reputations. We note that the reputations can be measured by interviews. Moreover, the aim of (Kulakowski and Gawronski 2009) has not been to compare the results with a given set of data, but rather to check if cooperation without payoffs is possible. No fitting has been done there. Similarly, the differential equations used to simulate the removal of cognitive dissonance in (Kulakowski et al. 2005) have been designed to illustrate the mechanism, and the accordance with experimental data of (Zachary 1977) should be treated as to some extent fortuitous. No parameters have been fitted there. Advantages of simple models are commonly known (Edmonds 2000; Edmonds 2005). Yet, as is also known, their flaw is that the condition of simplicity drives these models far from reality. We perceive their role to be similar to the one of verbal syllogisms in ancient philosophy: they should improve the clarity of our thinking of social systems.

The list of approaches presented above is limited to computational models, embedded in literature. It is worthwhile to mention also direct measurements of physiological variables which reflect emotional states of the subjects (Riener et al. 2009; Kashif et al. 2010). In some sense, however, the situation in these experiments mirrors the one in modeling. We expect that emotions are present in some situations, as when driving during the rush hours, we can even infer that a particularly risky strategy of driving, when observed, could be due to some specific mental state, but – more than often – the connection between behavior and emotions remains unverified. It is precisely the internal character of emotional states what makes the research of social systems so complex. We suspect that these states influence the system behavior and almost always we are right. But more insight into this internal world cannot be attained without a dedicated research. The next step – what determines these emotions? – is related with past experience of our subjects, and therefore it is even more far.

To summarize, either we can measure or at least evaluate our parameters, or the sensitivity analysis is necessary. In the case of internal parameters which can be measured only indirectly, the latter analysis seems unavoidable. Once stated, this rule seems trivial; yet sometimes the practice is different. In the social world of fluctuating parameters, an accordance of model predictions with a set of experimental data is often fortuitous. Therefore, on the contrary to natural sciences, it cannot be treated as the final proof of truth. The famous irony of John von Neumann "with four parameters I can fit an elephant, and with five I can make him wiggle his trunk" finds its target again.

## ACKNOWLEDGEMENTS

It is good occasion for one of authors (K.K.) to express his gratitude to Cristina Beltran-Ruiz, Vanessa Camilleri, Vikas Chandra, Maggie Ellis, Alois Ferscha, Matthew Fullerton, Przemek Gawronski, Jan Kantelhardt, Mirko Kampf, Ruediger Korff, Gosia Krawczyk, Paul Lukowicz, Krzysztof Malarz, Janusz Malinowski, Hermann de Meer, Matthew Montebello, Lev Muchnik, Eve Mitleton, Gevisa La Rocca, Khalid Saeed, Alexei Sharpanskykh, Wiesia Sikora, Jarek Wąs and Martin Wirz for kind discussions and cooperation. We are also grateful to our Referees for their criticism and tolerance.

**KRZYSZTOF KUŁAKOWSKI** was born in Zakopane, Poland. He studied theoretical physics at the Jagiellonian University in Cracow, and obtained MSc diploma in 1975. Since then he has been a member of staff of various faculties at the AGH University of Science and Technology in Cracow. He received PhD degree in solid state physics in 1984. He is a coauthor of 180 refereed papers. Since 2000 he is involved in applications of statistical physics to social phenomena. His e-mail address is: kulakowski@fis.agh.edu.pl and his Web-page can be found at: http://www.ftj.agh.edu.pl/~kulakowski/.

**PIOTR GRONEK** was born in Kielce, Poland. He studied technical physics at AGH University of Science and Technology (AGH-UST) in Cracow, and obtained MSc diploma in 1990. There, he received PhD degree in technical nuclear physics in 1997. Since then he has been a member of the teaching staff at Faculty of Physics and Applied Computer Science, AGH-UST. His e-mail address is: Piotr.Gronek@fis.agh.edu.pl and his Web-page can be found at: http://www.fis.agh.edu.pl/~gronek.

**ANTONI DYDEJCZYK** was born in Zabrze, Poland. He studied technical physics at AGH University of Science and Technology (AGH-UST) in Cracow, and obtained MSc diploma in 1981. Since then he has been a member of the staff at Faculty of Physics and Applied Computer Science, AGH-UST. He received PhD degree in technical nuclear physics in 2006. His e-mail address is: Antoni.Dydejczyk@fis.agh.edu.pl and his Web-page can be found at:
http://www.fis.agh.edu.pl/~antek.